\documentclass[amsmath,amssymb,showkeys,superscriptaddress,aps,prd,10pt,onecolumn]{revtex4-2}
\usepackage{graphicx}
\usepackage[utf8]{inputenc}
\usepackage[caption=false]{subfig}
\usepackage{multirow}
\usepackage{appendix}
\usepackage{graphicx,epstopdf}

\usepackage{colortbl}
\usepackage{xcolor}
\usepackage[colorlinks=true, urlcolor=blue, linkcolor=blue, citecolor=blue]{hyperref}
\usepackage{float}

\begin{document}

\title{Doubly-Charged $T_{cc}^{++}$ States in the Dynamical Diquark Model}
\author{Halil Mutuk}%
\email[]{hmutuk@omu.edu.tr}
\affiliation{Department of Physics, Faculty of Sciences, Ondokuz Mayis University, 55200, Samsun, Türkiye}

 
\begin{abstract}
One of the celebrated tools in explaining the Hydrogen atom is Born-Oppenheimer approximation. The resemblance of $QQ\bar{q}\bar{q}$ tetraquarks to Hydrogen atom within Quantum chromodynamics (QCD) implies usage of Born-Oppenheimer approximation for these multiquark states. In this work, we use dynamical diquark model to calculate mass spectra and sizes of doubly charmed and charged tetraquark states denoted as $T_{cc}^{++}$. Our results for mass spectra indicate some bound state candidates with respect to corresponding two-meson thresholds. Calculation of expectation values of $\sqrt{\langle r^2 \rangle}$ reflects that doubly charmed and charged tetraquark states are compact. 
\end{abstract}

\maketitle

\section{Prologue}\label{introduction}

It is well known that the existence of exotic states composed of more than three quarks or having unusual quantum numbers is not ruled out in QCD. Conventional hadrons can be grouped into mesons ($q\bar{q}$ states) and baryons ($qqq$ states). Exotic states in this perspective cannot fit into the conventional hadron configuration. Exotic particles have been observed in many experiments throughout the previous few decades. It is worth to mention that these exotic states were conjectured almost a half century ago phenomenologically \cite{GellMann:1964nj,Zweig:1981pd,Jaffe:1976ig,Jaffe:1976ih,Jaffe:1976yi}.

With the accumulation of experimental information about these exotic hadrons, many different models have been proposed to enlighten internal structure of these states \cite{Nielsen:2009uh,Chen:2016qju,Lebed:2016hpi,Ali:2017jda,Olsen:2017bmm,Guo:2017jvc,Brambilla:2019esw,Liu:2019zoy,Dong:2021juy}. Such endeavors are very valuable and important to understand perturbative and nonperturbative aspects of QCD. 

Recently, the LHCb Collaboration announced the first observation of doubly-charmed tetraquark state $T_{cc}^{+}$ in the $D^0 D^0 \pi^+$ invariant mass spectrum~\cite{LHCb:2021auc}. This state has a mass $M=3875.1 + \delta m_{exp}$, where $\delta m_{exp}=-273 \pm 61 \pm 5^{+11}_{-14} ~ \mathrm{keV} $ and width $\Gamma =410 \pm 165 \pm 43_{-38}^{+18}~\mathrm{keV}$ and has a quark content $cc \bar u \bar d$. In the unitarized Breit-Wigner analysis, the binding energy and decay width change to $\delta=360\pm 40^{+4}_{-0}$~keV and $\Gamma=48\pm 2^{+0}_{-14}$~keV, respectively~\cite{LHCb:2021vvq}. It is below two-meson $D^0  D(2010)^{\ast +}$ threshold and it is the longest-living exotic meson detected to date because of its very small width. The spin-parity quantum number of $T_{cc}^{+}$ state is determined as $J^P=1^+$ by the experiment. Due to the its interesting properties, the observation of $T_{cc}^{+}$ triggered many studies such as hadronic molecule generated  by the $DD^{*}$ interaction ~\cite{Du:2021zzh,Meng:2021jnw,Chen:2021vhg,Ling:2021bir,Feijoo:2021ppq,Yan:2021wdl,Ke:2021rxd,Dong:2021bvy,Huang:2021urd,Fleming:2021wmk,Hu:2021gdg,Chen:2021cfl,Deng:2021gnb,Agaev:2022ast,He:2022rta,Abreu:2022lfy,Wang:2022jop,Wang:2023ovj}, compact tetraquark \cite{Agaev:2021vur,Azizi:2021aib, Abreu:2022lfy}, or triangle singularity \cite{Braaten:2022elw}. 

Heavy quark spin symmetry (HQSS) \cite{Neubert:1993mb} suggests that when the heavy quark masses are taken to infinity $m_Q \to \infty$, the spin of the quark decouples from the dynamics which refers the strong interactions in the system are independent of the heavy quark spin. HQSS predicts the presence of partners for the particles involving heavy quarks. For example using HQSS, if $\chi_{c1}(3872)$ (formerly known as $X(3872)$) is a bound state of $D$ and $D^\ast$ mesons, then it should have partners in heavy quark limit \cite{Nieves:2012tt,Hidalgo-Duque:2012rqv}. In the same manner, locations of possible $T_{cc}^{++}$ and $T_{cc}^0$ states were determined in Ref. \cite{Albaladejo:2021vln}. In the same work, $I(J^P)=0(1^+)$ and $I(J^P)=1(2^+)$ HQSS partners below $D^{\ast +} D^{\ast 0}$ thresholds were predicted. In Ref. \cite{Agaev:2018vag}, masses and couplings of the $T_{cc}$ states with light diquark content $\bar{s}\bar{s}$ and $\bar{d}\bar{s}$ were studied in QCD sum rule method. Scalar, axialvector and tensor doubly-charmed molecular four-quark states with isospin $I=0,1/2$ and $I=1$ have been studied in QCD sum rule framework \cite{Xin:2021wcr}. Alongside $T_{cc}^+$ state, masses of $T_{cc}$ states with single and double strangeness have been examined using one-meson exchange potential model \cite{Ren:2021dsi}. In Ref. \cite{Ozdem:2021hmk}, magnetic moments of hypothetical $T_{cc}^{++}$ states were calculated in molecular picture using light-cone QCD sum rule method. Ref. \cite{Andreev:2021eyj} studied $QQ \bar{q} \bar{q}$ system in the Born-Oppenheimer approximation within string theory.

It is well known that doubly charged four-quark states are pure exotic states since they cannot be explained as traditional mesons. The most recent example is the observation of a doubly charged open-charm tetraquark state with possible quark content $(c \bar{s} u \bar{d})$ by LHCb Collaboration in 2022 \cite{LHCb:2022sfr}. Doubly charmed four-quark states can be interpreted as compact tetraquark states or molecular states. A recent experimental observation in radiative decays of $\chi_{c1}(3872)$ provides a strong argument in favour of a compact component in the $\chi_{c1}(3872)$ structure \cite{LHCb:2024tpv}. This state  is just a few keV below corresponding $D^0 \bar{D}^{*0}$ threshold and the ubiquitous point of view is that $\chi_{c1}(3872)$ state has a molecular nature. In addition to this, there is research on the system formed mainly due to the electric Coulomb force, $D^{\pm}D^{*\mp}$ \cite{Zhang:2024fxy} which can be used as a key to reveal inner structure of $\chi_{c1}(3872)$. Hadronic molecules are extended objects. For example, as stated in Ref. \cite{Guo:2017jvc}, if $\chi_{c1}(3872)$ has a molecular nature it would be at least as large as 10 fm. If doubly-charmed tetraquarks have double electric charge, these states may not be interpreted as loosely bound molecules even if they occur at threshold, because of the Coulomb barrier.  In a loosely bound molecule, an extended object with respect to the range of strong interactions, Coulomb repulsion would induce a fall-apart decay. Therefore doubly-charmed and doubly-charged tetraquarks may be interpreted as compact tetraquarks even one of them was experimentally observed close to any open charm threshold \cite{Esposito:2013fma}.

To the best of our knowledge, there is no study about doubly charmed and charged states conducted in dynamical diquark model. In  this paper, we present mass spectra and sizes of doubly charmed and charged tetraquark states using dynamical diquark model. In Section \ref{formalism}, we introduce dynamical diquark model and in Section \ref{Potential} we give brief information about Born-Oppenheimer approximation and potential. Section \ref{numerical} presents results of numerical calculations and Section \ref{final} summarizes our results.

\section{Dynamical Diquark Model} \label{formalism}
The resolution of a compact tetraquark can be written as a diquark+antidiquark state in which diquark (antidiquark) is treated as a degrees-of-freedom.  A diquark is a colored bound quark-quark pair $(qq)$ and antidiquark is a colored bound antiquark-antiquark pair $(\bar q \bar q)$. These colored pairs can have colorless combinations. A consideration of the structure and spectroscopy of multiquark states may even make diquark degrees-of-freedom important. For more discussion about diquarks see reviews \cite{Anselmino:1992vg,Barabanov:2020jvn}. 

The tetraquark interpretation of a four-quark state can be done by using diquarks and antidiquarks as $[(qq)(\bar{q}\bar{q})]$ combination. In this configuration, the quark $q$ is in the fundamental color representation 3 and the antiquark $\bar{q}$ is in the fundamental color representation $\bar{3}$. Combining two quarks as $\vert qq \rangle$ gives $3 \otimes 3= 6 \oplus \bar{3} $, whereas combining two antiquarks $\vert \bar	q \bar q \rangle$ gives $\bar 3 \otimes \bar 3= \bar 6 \oplus 3$. Accordingly, color singlet tetraquark $\vert qq \bar q \bar q \rangle$ state may be produced from $\bar 3 \otimes 3$ or $\bar 6 \times 6$.

The dynamical diquark model proposed in Ref. \cite{Brodsky:2014xia} suggests that confinement mechanism is responsible for binding in the exotic states. The diquark-antidiquark pair forms instantly at the production point and rapidly moves from each other due to the kinematic effects of the production process. Due to the fact that diquark and antidiquark are colored objects, they cannot separate far apart; instead they elongate creating a {\it color flux tube\/} or {\it string\/} between them. This model was developed and upgraded in Refs. \cite{Giron:2019bcs,Giron:2019cfc} where diquarks $\delta$ and antidiquarks $\bar{\delta}$ are treated as quasi-bound hadronic subcomponents.

The states for the $\delta-\bar{\delta}$ system can be written 
\begin{eqnarray}
J^{PC} = 0^{++}: & \ & X_0 \equiv \vert 0_{\delta} , 0_{\bar{\delta}}\rangle_0 \, , \ \ X_0^\prime \equiv \vert 1_{\delta} , 1_{\bar{\delta}} \rangle_0, \nonumber \\
J^{PC} = 1^{++}: & \ & X_1 \equiv \frac{1}{\sqrt 2} \left(
\vert 1_{\delta} , 0_{\bar{\delta}}\rangle_1  + \vert 0_{\delta} , 1_{\bar{\delta}}\rangle_1 \right) , \nonumber \\
J^{PC} = 1^{+-}: & \ & Z \ \equiv
\frac{1}{\sqrt 2} \left(
\vert 1_{\delta} , 0_{\bar{\delta}}\rangle_1  - \vert 0_{\delta} , 1_{\bar{\delta}}\rangle_1 \right) , \nonumber \\ & \ & Z^\prime \,
\equiv \vert 1_{\delta} , 1_{\bar{\delta}} \rangle_1 \, , \nonumber \\
J^{PC} = 2^{++}: & \ & X_2 \equiv \vert 1_{\delta} , 1_{\bar{\delta}} \rangle_2 \ ,
\label{diquarkstates}
\end{eqnarray}
where the basis of good diquark-spin eigenvalues with labels such as $1_{\delta}$ are used. Here outer subscripts on the kets indicate total quark spin $J$. Since we study $S$-wave states, it equals $J=S$. These six states 
constitute the lowest multiplet $\Sigma^+_g(1S)$ within the Born-Oppenheimer  approximation. The core states $X_0$, $X_0^\prime$ and $Z$, $Z^\prime$ have the same $J^{PC}$ quantum numbers and can mix. Instead  equivalent eigenstates \cite{Giron:2020qpb} can be defined as $X_1$, $X_2$, and 
\begin{eqnarray}
{\tilde X}_0 & \equiv & \vert 0_{Q\bar{Q}}
 , 0_{q^\prime \bar{q}^\prime} \rangle_0 =
+ \frac{1}{2} X_0 + \frac{\sqrt{3}}{2} X_0^\prime \, , \nonumber \\
{\tilde X}_0^\prime & \equiv & \vert 1_{Q\bar{Q}} , 1_{q^\prime \bar{q}^\prime} \rangle_0 =
+ \frac{\sqrt{3}}{2} X_0 - \frac{1}{2} X_0^\prime \, , \nonumber \\
{\tilde Z} & \equiv & \vert 1_{Q\bar{Q}} , 0_{q^\prime \bar{q}^\prime} \rangle_1 =
\frac{1}{\sqrt{2}} \left( Z^\prime \! + Z \right) \, , \nonumber \\
{\tilde Z}^\prime & \equiv & \vert 0_{Q\bar{Q}} , 1_{q^\prime \bar{q}^\prime} \rangle_1 =
\frac{1}{\sqrt{2}} \left( Z^\prime \! - Z \right) \, .
\label{diquarkstatesmix}
\end{eqnarray}

If the quarks inside of $\delta$ or $\bar{\delta}$ pairs are identical, then the space-spin wave function of the related diquark must be symmetric. As a result, the total wave function together with color part of the corresponding diquark obeys the Fermi statistic. The dynamical diquark model assumes no orbital excitation within the diquarks. Therefore spatial wave functions and spin wave functions of diquarks should be symmetric which leaves only $1_ {\delta}$ and $1_{\bar{\delta}}$  survive \cite{Giron:2020wpx}. This is the case for $cc$, $\bar{d}\bar{d}$, and $\bar{s} \bar{s}$ diquarks but not in the $\bar{d} \bar{s}$ which can combine together to have $0_{\bar{\delta}}$.  This elaboration requires more involved analysis which is beyond the assumption of this work. It is assumed that the tetraquark states in this present study are composed of colour antitriple ($\bar{3}_c$) diquark  and colour triplet ($3_c$) antidiquark. Therefore $X_0^\prime$, $Z^\prime$ and $X_2$ are the eigenstates that remain and no mixing occurs. Furthermore, the states that are considered is this work do not have a specified $C$- parity. As a result for these states, it is not a "good" quantum number.

\section{Born-Oppenheimer Approximation and Potential}\label{Potential} 
The original Born-Oppenheimer (BO) approximation was proposed to study quantum states of molecules made of nuclei and electrons. In this approximation, it is possible to separate the motion of the nuclei and the motion of the electrons, i.e., the separation of nuclear and electronic coordinates facilitates to solve Schrödinger equation for multibody systems. The nuclei is sufficiently heavy and its motion is approximately frozen where the electrons around the nuclei move very much faster. The nuclei constitute a big attractive force such that electrons cannot ``disappear" from the atom.

From the four-quark point of view, at least two heavy sources and degrees-of-freedom for light quarks are needed for the Born-Oppenheimer approximation. In the doubly-heavy $QQ \bar q \bar q$ tetraquarks heavy quark mass is much bigger than the light quark mass, $M_Q >> m_q$. In this limit, fast motion of light quarks in the presence of the heavy color sources produces an effective potential. This potential regulates the slow motion of heavy quarks. The slow motion of  heavy quarks falls within the so-called Born-Oppenheimer potential. \cite{Maiani:2022qze}.

In this work, mass spectrum of the $T_{cc}^{++}$ states is computed by a specific BO potential, $\Gamma$ ($\Sigma_g^+$, $\Sigma_u^+$, \emph{etc}.) which gives rise to a multiplet of states. These potentials are obtained by several groups on lattice \cite{Juge:1997nc,Juge:1999ie,Juge:2002br,Morningstar:2019,Capitani:2018rox}. Having masses of a diquark $m_{\delta}$ and antidiquark $m_{\bar{\delta}}$, one can solve numerically the corresponding Schr\"{o}dinger equation. 

The Born-Oppenheimer approximation in QCD has a numerically accessible potential obtained by lattice QCD (see for example Refs. \cite{Juge:1999ie,Bicudo:2015kna}) which describe successfully heavy quark-antiquark systems. Refs. \cite{Brown:2012tm,Bicudo:2012qt,Bicudo:2016ooe,Bicudo:2017szl,Pflaumer:2018hmo,Prelovsek:2019ywc} used Born-Oppenheimer approximation using lattice QCD potentials. Due to this feature, Born-Oppenheimer approximation prevails being one of the promising tools for studying both conventional and unconventional heavy quark-antiquark states \cite{Braaten:2014qka,Brambilla:2017uyf,Bruschini:2023zkb}.

The radial Schr\"{o}dinger equations for the uncoupled BO potentials $V_\Gamma(r)$ can be written as, in natural units for reduced mass $\mu$, 
\begin{equation}
\left[-\frac{1}{2\mu r^2}\partial_r r^2 \partial_r + \frac{\ell\left(\ell +1\right)}{2\mu r^2}+V_\Gamma(r)\right]\psi_\Gamma^{(n)}(r)=E_n\psi_\Gamma^{(n)},
\end{equation}
and the coupled BO potentials read:
\begin{eqnarray}
&&\left[-\frac{1}{2\mu r^2}\partial_r r^2 \partial_r +\frac{1}{2\mu r^2}
\begin{pmatrix}
\ell\left(\ell+1\right) +2 & 2\sqrt{\ell\left(\ell+1\right)}\\
2\sqrt{\ell\left(\ell+1\right)} & \ell\left(\ell+1\right)
\end{pmatrix}\right.\nonumber\\
&& \left. 
\begin{pmatrix}
V_{\Sigma_u^-} & 0\\
0 & V_{\Pi_u^+}
\end{pmatrix}\right]
\begin{pmatrix}
\psi_{\Sigma_u^-}^{(n)}(r)\\
\psi_{\Pi_u^+}^{(n)}(r)
\end{pmatrix}= E_n
\begin{pmatrix}
\psi_{\Sigma_u^-}^{(n)}(r)\\
\psi_{\Pi_u^+}^{(n)}(r)
\end{pmatrix}.
\end{eqnarray}

We use the functional form of $V(r)$ given by lattice simulations \cite{Juge:2002br,Morningstar:2019}. The $S$-wave Hamiltonian can be written as
\begin{eqnarray} 
H&=& H_0 + 2\left[\kappa_{QQ}\left(\bold{s}_Q \cdot \bold{s}_Q \right)+
\kappa_{\bar{q}\bar{q}} \left(\bold{s}_{\bar{q}} \cdot \bold{s}_{\bar{q}}\right)
\right].
\label{Hamiltonian}
\end{eqnarray}
The eigenvalues of $H$ are computed as
\begin{equation}
M= M_0 + \Delta M_{\kappa_{QQ}} + \Delta M_{\kappa_{\bar{q}\bar{q}}},
\end{equation}
where 
\begin{equation}
\Delta M_{\kappa_{QQ}} =\frac 1 2 \kappa_{QQ}\left[2s_\delta \left(s_\delta +1\right)-3
\right],
\end{equation}
and
\begin{equation}
\Delta M_{\kappa_{\bar{q}\bar{q}}}= \frac 1 2 \kappa_{\bar{q}\bar{q}}
\left[2s_{\bar{\delta}} \left(s_{\bar{\delta}} +1\right) -3 \right].
\end{equation}
In the above expressions, $s_\delta$ represents the spin of the heavy diquark and $s_{\bar{\delta}}$ represents the spin of the light antidiquark. 

The Hamiltonian in Eq. (\ref{Hamiltonian}) first appeared in Ref. \cite{Lebed:2016yvr} and also included orbital and spin-orbit terms. Ref. \cite{Giron:2019cfc} used S-wave Hamiltonian to study $c \bar{c} q \bar{q}^{\prime}$ states. The same Hamiltonian emerged in Ref. \cite{Giron:2020qpb} for hidden-bottom and hidden-charm-strange states and in Ref. \cite{Giron:2020wpx} for all-heavy tetraquark states. The authors obtained reliable results compared to literature. 

In the model of this present work, we assume that there is no spin-spin interaction between $\delta$ and $\bar{\delta}$ pairs. This assumption is supported by the conclusion that the dominant spin-spin interactions in the $\delta- \bar{\delta}$ states appear to be the ones within each of $\delta$ and $\bar{\delta}$ diquarks \cite{Maiani:2014aja}.

\section{Mass Spectrum}\label{numerical}

The possible quark content of doubly charmed  and charged tetraquark states are $cc \bar d \bar d$, $cc \bar d \bar s$, and  $cc \bar s \bar s$. We assume that diquarks are pointlike objects. Diquark and antidiquark masses $(m_{\delta},m_{\bar{\delta}})$  are need to be adjusted to the model as input. In order to make predictions that do not heavily depend on nonrelativistic quark models \cite{Lebed:2017min,Mutuk:2022nkw}, we use  diquark mass values that are extracted from QCD sum rule $M_{ud}(1^+)=0.81 \pm 0.06 ~\text{GeV}$ and $M_{qs}(1^+)=0.92 \pm 0.04 ~\text{GeV}$ from Ref. \cite{Wang:2011ab}, and $M_{cc}(1^+)=3.51 \pm 0.35 ~\text{GeV}$ from Ref. \cite{Esau:2019hqw}. A recent study conducted in QCD sum rule for $J^P=1^+$ $ss$ diquark is failed to stabilize masses \cite{deOliveira:2023hma}. Therefore we use doubly-strange diquark mass from Ref. \cite{Ebert:2011kk} as $M=1203~ \text{MeV}$. The coupling constants are taken as $\kappa_{qq}=\frac{1}{2}\kappa_{q\bar{q}}$ (which is a general consideration for the diquark potential is the half of the corresponding quark-antiquark potential) where $\kappa_{q\bar{q}}=315 ~ \text{MeV}$, $\kappa_{s\bar{q}}=195 ~ \text{MeV}$, and $\kappa_{s\bar{s}}=121 ~ \text{MeV}$ \cite{Maiani:2004vq}. 

\subsection{$cc \bar d \bar d$ State}

The mass spectrum and expectation values for $\sqrt{\langle r^2 \rangle}$ of $cc \bar d \bar d$ state are given in Table \ref{tab:ccdd1}. We also show values of the difference of the doubly-heavy tetraquark and threshold masses, $\Delta=M-T$ in the same table. A negative $\Delta$ means the state is below corresponding two-meson threshold and could be observed as a bound state. Positive but small values of $\Delta$ correspond to resonance states which can be observed as resonances. Remaining values of $\Delta$ reflect that states should be broad and are difficult to observe experimentally. We also give the uncertainties in the mass values due to the lattice QCD inputs. Since the uncertainties are less than $\% 1$ in the $\sqrt{\langle r^2 \rangle}$ values, we refrain to present their values in all tetraquark cases of this work. As can be seen from the Table, mass splitting of $cc \bar d \bar d$ multiplet is 97 MeV. We observe from $\sqrt{\langle r^2 \rangle}$ values that the predicted states are compact.

\renewcommand{\tabcolsep}{0.60cm} \renewcommand{\arraystretch}{1.1}
\begin{table}[H]
\centering
\caption{Predicted mass spectra, $\sqrt{\langle r^2 \rangle}$, corresponding $\Delta$ values for thresholds of $cc \bar d \bar d$ state. The superscripts denote the spin of the diquark/antidiquark whereas the subscripts denote the color structure. $\{~\}$ denotes the symmetric flavor wave functions of the two quarks and antiquarks subsystems. Mass results are in MeV and $\sqrt{\langle r^2 \rangle}$ values are in fm units.}
\label{tab:ccdd1}%
\begin{tabular}{ccccccc}
\hline\hline
 $J^{P}$ & Configuration  & Mass & $\sqrt{\langle r^2 \rangle}$   &Threshold &$\Delta$  \\ \hline
$0^{+}$  & $\left[\{cc\}^1_{\bar{\mathbf{3}}_c}\{\bar{d}\bar{d}\}^1_{\mathbf{3}_c}\right]^0_{\mathbf{1}_c}$  & $3904 \pm 76$ & 0.64 &$DD$ & 170 \\
$1^{+}$  & $\left[\{cc\}^1_{\bar{\mathbf{3}}_c}\{\bar{d}\bar{d}\}^1_{\mathbf{3}_c}\right]^1_{\mathbf{1}_c}$  & $ 3952 \pm 80$ & 0.65&$DD^\ast$ & 76  \\
$2^{+}$  & $\left[\{cc\}^1_{\bar{\mathbf{3}}_c}\{\bar{d}\bar{d}\}^1_{\mathbf{3}_c}\right]^2_{\mathbf{1}_c}$  & $4001 \pm 82 $  & 0.65&$D^\ast D^\ast$  & -16 \\\hline
 \hline
\end{tabular}%
\end{table}

It can be seen from Table \ref{tab:ccdd1} that the masses of $J^P=0^+$ and $J^P=1^+$ $cc \bar d \bar d$ states are above corresponding thresholds. They could be resonance states. The mass of $J^P=2^+$ $cc \bar d \bar d$ state is 16 MeV below corresponding threshold. This state could be a bound state candidate.

\subsection{$cc \bar d \bar s$ State}

The mass spectrum, expectation values for $\sqrt{\langle r^2 \rangle}$ and values of the $\Delta$ for the $cc \bar d \bar s$ tetraquarks are presented in Table \ref{tab:ccds1}. As can be seen from the table, mass splitting of $cc \bar d \bar s$ multiplet is 87 MeV. We observe from $\sqrt{\langle r^2 \rangle}$ values that the predicted states are compact. Masses of $J^P=0^+$ and $J^P=1^+$ $cc \bar d \bar s$ states are above corresponding thresholds. They could be resonance states. The mass of $J^P=2^+$ $cc \bar d \bar s$ state is 52 MeV below corresponding threshold. This state could be a bound state candidate. 

\renewcommand{\tabcolsep}{0.60cm} \renewcommand{\arraystretch}{1.1}
\begin{table}[H]
\centering
\caption{Same as Table \ref{tab:ccdd1} but for $cc \bar d \bar s$ state.}
\label{tab:ccds1}%
\begin{tabular}{ccccccc}
\hline\hline
 $J^{P}$ & Configuration & Mass & $\sqrt{\langle r^2 \rangle}$  & Threshold &$\Delta$  \\ \hline
$0^{+}$ & $\left[\{cc\}^1_{\bar{\mathbf{3}}_c}\{\bar{d}\bar{s}\}^1_{\mathbf{3}_c}\right]^0_{\mathbf{1}_c}$ & $3982 \pm 86$ & 0.64& $DD_s$ & 146 \\
$1^{+}$ & $\left[\{cc\}^1_{\bar{\mathbf{3}}_c}\{\bar{d}\bar{s}\}^1_{\mathbf{3}_c}\right]^1_{\mathbf{1}_c}$ & $4031 \pm 90$ & 0.65& $D^\ast D_s/D_s^\ast D$ & 54/52  \\
$2^{+}$ & $\left[\{cc\}^1_{\bar{\mathbf{3}}_c}\{\bar{d}\bar{s}\}^1_{\mathbf{3}_c}\right]^2_{\mathbf{1}_c}$  & $4069 \pm 93$  & 0.65&$D^\ast D_s^ \ast$  & -52 \\\hline
 \hline
\end{tabular}%
\end{table}

\subsection{$cc \bar s \bar s$ State}
The mass spectrum, expectation values  for $\sqrt{\langle r^2 \rangle}$,  and values of the $\Delta$  for $\sqrt{\langle r^2 \rangle}$ of $cc \bar s \bar s$ state are given in Table \ref{tab:ccss1}. As can be seen from the table, mass splitting of $cc \bar d \bar s$ multiplet is 57 MeV. We observe from $\sqrt{\langle r^2 \rangle}$ values that the predicted states are compact. masses of $J^P=0^+$ and $J^P=1^+$ $cc \bar s \bar s$ states are above corresponding thresholds. They could be resonance states. The mass of $J^P=2^+$ $cc \bar s \bar s$ state is 46 MeV below corresponding threshold. This state could be a bound state candidate.

\renewcommand{\tabcolsep}{0.60cm} \renewcommand{\arraystretch}{1.1}
\begin{table}[H]
\centering
\caption{Same as Table \ref{tab:ccdd1} but for $cc \bar s \bar s$ state.}
\label{tab:ccss1}%
\begin{tabular}{cccccc}
\hline\hline
 $J^{P}$ & Configuration &  Mass &   $\sqrt{\langle r^2 \rangle}$  & Threshold & $\Delta$  \\ \hline
$0^{+}$ & $\left[\{cc\}^1_{\bar{\mathbf{3}}_c}\{\bar{s}\bar{s}\}^1_{\mathbf{3}_c}\right]^0_{\mathbf{1}_c}$ & $4121 \pm 104$  & 0.64 &$D_sD_s$ & 184 \\
$1^{+}$ & $\left[\{cc\}^1_{\bar{\mathbf{3}}_c}\{\bar{s}\bar{s}\}^1_{\mathbf{3}_c}\right]^1_{\mathbf{1}_c}$ & $4148 \pm 105$ & 0.65& $D_s D_s^\ast$ & 67  \\
$2^{+}$ & $\left[\{cc\}^1_{\bar{\mathbf{3}}_c}\{\bar{s}\bar{s}\}^1_{\mathbf{3}_c}\right]^2_{\mathbf{1}_c}$ & $ 4178 \pm 112$ & 0.65& $D^\ast D_s^ \ast$  & -46 \\\hline
 \hline
\end{tabular}%
\end{table}

\subsection{Comparison with other works}
In Table \ref{tab:comparison}, we confront our results for masses of doubly-charmed tetraquark states with other theoretical studies \cite{Ebert:2007rn,Wang:2017dtg}. In Ref. \cite{Ebert:2007rn}, within the context of the diquark-antidiquark picture in the relativistic quark model, masses of tetraquarks with two heavy quarks and open charm and bottom are computed. In Ref. \cite{Wang:2017dtg}, doubly charmed tetraquark states are studied with QCD sum rules formalism. As can be seen in the related table,  our results agree well with the results of Ref. \cite{Wang:2017dtg} in which QCD sum rules formalism is used. The authors constructed the axialvector-diquark-axialvector-antidiquark type currents to study scalar, axialvector, vector, tensor doubly charmed tetraquark states. In the $cc \bar{d} \bar{d}$ tetraquarks, our central mass values are approximately 100 MeV lower than the results of Ref. \cite{Ebert:2007rn}, whereas in the $cc \bar{q} \bar{s}$ and $cc \bar{s} \bar{s}$ tetraquarks, our results are approximately 200 MeV lower. The reason for this could be that different models, interactions and parameters are used in both works. 

\renewcommand{\tabcolsep}{0.60cm} \renewcommand{\arraystretch}{1.1}
\begin{table}[H]
\centering
\caption{Comparison of different theoretical predictions for the doubly-charmed tetraquark states.}
\label{tab:comparison}%
\begin{tabular}{ccccc}
\hline\hline
 $J^{P}$ & Configuration &  This work &  \cite{Ebert:2007rn} &\cite{Wang:2017dtg}  \\ \hline
$0^{+}$ & $cc \bar{q} \bar{d}$ &  $3904 \pm 76 ~ \mbox{MeV}$  & $4056 ~ \mbox{MeV}$ & $3.87 \pm 0.09 ~ \mbox{GeV}$ \\
$1^{+}$ & & $3952 \pm 80$ & 4079  &  $3.90 \pm 0.09$   \\
$2^{+}$ &  & $4001 \pm 82$ & 4118  & $3.95 \pm 0.09$  \\\hline

$0^{+}$ & $cc \bar{q} \bar{s}$ & $3982 \pm 86$ & 4221 & $3.94 \pm 0.10$ \\
$1^{+}$ & & $4031 \pm 90$  & 4239  &  $3.96 \pm 0.08$   \\
$2^{+}$ &  &  $4069 \pm 93$ & 4271  & $4.01 \pm 0.09$  \\\hline

$0^{+}$ & $cc \bar{s} \bar{s}$ & $4121 \pm 104$ & 4359 & $3.99 \pm 0.10$ \\
$1^{+}$ & & $4148 \pm 105$   & 4375  &  $4.02 \pm 0.09$   \\
$2^{+}$ &  &  $ 4178 \pm 112$ & 4402  & $4.06 \pm 0.09$  \\\hline

 \hline
\end{tabular}%
\end{table}

\section{Discussion and Final Remarks}\label{final}

We obtain mass spectra and expectation values of $\sqrt{\langle r^2 \rangle}$ for the $cc \bar d \bar d$, $cc \bar d \bar s$, and  $cc \bar s \bar s$ tetraquark states using dynamical diquark model. The calculated $\sqrt{\langle r^2 \rangle}$ values reflect the distance between heavy diquark and light antidiquark. As can be seen from Tables \ref{tab:ccdd1}, \ref{tab:ccds1}, and \ref{tab:ccss1} we obtain some bound state candidates. A calculation of strong decays for the $cc \bar d \bar d$, $cc \bar d \bar s$, and  $cc \bar s \bar s$ tetraquark states can give information about existence of bound states. From the Tables \ref{tab:ccdd}, \ref{tab:ccds}, and \ref{tab:ccss} it can be seen that the $cc \bar d \bar d$, $cc \bar d \bar s$, and  $cc \bar s \bar s$ tetraquark states are compact. The size of the tetraquarks is less than $< 1~ \text{fm}$ which is called confinement radius. 

The observation of $T_{cc}^+$ state is as important as the observation of $\chi_{c1}(3872)$ (formerly known as X(3872)). The possible observation of $T_{cc}^{++}$ states in future experimental studies would constitute a new establishment for multiquark states. We hope that our results will help for this endeavour.

\bibliography{Tccplusplus-revised2}

\end{document}